\documentclass[letterpaper, 10pt, conference]{ieeeconf}
\IEEEoverridecommandlockouts
\usepackage{amsmath,amsfonts,bm}
\usepackage{graphicx}
\usepackage{subfigure}

\title{\LARGE \bf
Loop Calculus Helps to Improve Belief Propagation and Linear
Programming Decodings of Low-Density-Parity-Check Codes}

\author{Michael Chertkov and Vladimir Y. Chernyak
 \thanks{Invited talk at $44$-th annual Allerton Conference on
 Communications, Control and Computing, Sep 27-Sep 29, 2006.}
 \thanks{We are thankful to M. Stepanov for many great
 remarks and discussions. This work was carried out under the
auspices of the National
 Nuclear Security Administration of the
 U.S. Department of Energy at Los Alamos National Laboratory under
 Contract No. DE-AC52-06NA25396. VYC also acknowledges the support of WSU}
\thanks{M.~Chertkov is with Theoretical Division and Center for
Nonlinear Studies, LANL, Los Alamos, NM 87545, USA;
{\tt\small chertkov@lanl.gov}}%
\thanks{V.Y.~Chernyak is with  Department of Chemistry
Wayne State University 5101 Cass Ave Detroit, MI 48202;
{\tt\small chernyak@chem.wayne.edu}}%
}

\begin{document}

\maketitle
\thispagestyle{empty}
\pagestyle{empty}

\begin{abstract}
We illustrate the utility of the recently developed loop calculus
\cite{06CCa,06CCb} for improving the Belief Propagation (BP)
algorithm. If the algorithm that minimizes the Bethe free energy
fails we modify the free energy by accounting for a critical loop in
a graphical representation of the code. The log-likelihood specific
critical loop is found by means of the loop calculus. The general
method is tested using an example of the Linear Programming (LP)
decoding, that can be viewed as a special limit of the BP decoding.
Considering the $(155,64,20)$ code that performs over
Additive-White-Gaussian channel we show that the loop calculus
improves the LP decoding and corrects all previously found dangerous
configurations of log-likelihoods related to pseudo-codewords with
low effective distance, thus reducing the code's error-floor.
\end{abstract}

Belief Propagation (BP) constitutes an efficient approximation, as
well as an algorithm, that applies to many inference problems in
statistical physics \cite{35Bet,36Pei,82Bax}, information theory
\cite{63Gal,68Gal,99Mac,03RU}, and computer science \cite{88Pea}.
All these problems can be stated in terms of computation of marginal
probabilities on a factor graph. If the underlying graph structure
contains no loops, i.e. it is a tree, the BP is exact, being only an
approximation in the case of a general graph. The BP approximation
can be restated in terms of a variational principle
\cite{51Kik,91Mor,05YFW},  where the BP equations describe a minimum
of the so-called Bethe free energy, and the standard BP algorithm
\cite{63Gal,68Gal} means solving the BP equations iteratively.

In coding theory BP plays a special role as the decoding of choice
for Low Density Parity Check (LDPC) Codes introduced by R. Gallager
\cite{63Gal}. These codes, described in terms of sparse (Tanner)
graphs, are among the best performing codes known to date. Actually,
these codes perform so well exactly due to the high-quality
performance of the computationally efficient BP decoding scheme
\cite{63Gal,99Mac,03RU}. In the water-fall domain, i.e. at low and
moderate Signal-to-Noise Ratios (SNR), the Frame-Error-Rate (FER) or
Bit-Error-Rate (BER) of an LDPC code decoded using BP comes close to
the optimal yet inefficient Maximum-Likelihood and
Maximum-a-Posteriori decodings. However, in the low noise regime BP
decoding clearly fails to approximate ML in practical (finite size)
LDPC codes, thus causing the error-floor \cite{01MB,03TJVW,03Ric}.
It is now well understood that the BP decoding failure in the
error-floor domain is due to existence of pseudo-codewords
\cite{96Wib,98AHMX,99FKKR,01FKV,03KV,05KLVW} and related instantons
\cite{05SCCV,06SCa}, defined as dangerous noise configurations
causing the failures. Removing the obstacles and thus improving the
BP decoding while keeping a reasonable computational complexity is
on a great demand for high-performance applications, e.g. optical
communications and data storage, where error-floor is a serious
handicap.

Some general purpose BP-improvement strategies were already
discussed in the literature \cite{05YFW,02MPZ,05MR}. Survey
Propagation (SP) has been suggested in the context of combinatorial
optimization \cite{02MPZ}. SP, based on the so-called replica
approach that originates from spin glass theory \cite{87MPV},
applies successfully to highly degenerate problems where the
standard BP would be trapped in a local unrepresentative minimum.
Generalized Belief Propagation (GBP) of \cite{05YFW} constitutes
another generalization of BP. It extends the cluster variational
approximation of statistical physics \cite{51Kik,91Mor} to problems
in information and computer sciences. GBP has been shown to perform
well for the problems with many short loops, like Inter-Symbol
Interference on a regular two-dimensional lattice \cite{04SWSW},
where transition to a coarse-grained cluster offers an improvement
over the standard BP. The only yet serious drawback of the method is
the expense of an overhead that scales exponentially with the
cluster size. An efficient alternative to GBP, that is claimed
successful in describing some random graph and random spin models on
lattices, has been recently discussed in \cite{05MR}. The method is
based on closing the system of cavity equations and suggests a way
to account for correlations (many loops) on all scales.

In spite of their potential in dealing with highly degenerate
problems where the bare BP approach fails, the set of the
aforementioned methods/algorithms (maybe except for a possible
extension of \cite{05MR}) do not seem appropriate in dealing with
the error-floor problem. Indeed, one expects that any actual code
(as opposed to ensemble of codes) has a discrete set of well-defined
and relatively simple troublemakers (pseudo-codewords) localized on
a subgraph. Therefore, for each (rare) failure of the standard BP
one needs to identify and correct for a relatively long correlated
configuration that is localized however on a small portion of the
total number of bits.

This paper suggests an efficient approach that generalizes the BP
algorithm and which is capable of reducing the undesirable error
floor effect. Our method is based on the recently developed
analytical tool, called loop calculus \cite{06CCa,06CCb}, that
represents the partition function (and, therefore, the marginal
probabilities) in terms of a finite series where each term is
associated with a generalized loop on the graph and the zero order
contribution corresponds to the bare BP approximation. We conjecture
that for an instanton noise configuration that causes a BP failure
there is always a relatively simple loop correction to the bare BP
(in the terminology of the loop calculus) that provides an equal or
comparable contribution to the partition function. We further
suggest an improved decoding scheme based on finding this critical
loop correction (in the case of the bare BP failure) followed by
correcting the error. This is achieved by a proper modification of
the Bethe free energy and the BP equations. These ideas are verified
using an example of the Tanner $(155,64,20)$ code \cite{01TSF}
performing over the Additive-White-Gaussian-Noise (AWGN) channel and
decoded with the Linear Programming (LP) decoding \cite{03FWK}. We
build this test on analysis of the set of pseudo-codewords recently
found for this code by the LP-based pseudo-codeword search algorithm
\cite{06CS}. We introduce the LP-erasure decoding which is
equivalent to the standard LP decoding with full or partial erasure
of information at the bits along the critical loop. We demonstrate
that the LP-erasure algorithm corrects errors associated with all
previously found pseudo-codewords of the Tanner code ($\sim 200$ of
them) completely closing the error-floor gap between the lowest
LP-instanton, with the effective distance $\approx 16.4037$, and the
Hamming distance $20$ of the code.

The manuscript is organized as follows. An extended introductory
Section \ref{sec:Intro} consists of four Subsections. Subsection
\ref{subsec:SI} introduces the notations and briefly overviews
decoding of a binary linear code. Subsection \ref{subsec:LC}
describes the loop calculus of \cite{06CCa,06CCb} also complementing
it by some new variational interpretation that was not discussed in
the original papers. Subsection \ref{subsec:mag} describes the
calculation of a-posteriori log-likelihoods (magnetizations) within
the loop calculus. Subsection \ref{subsec:BFE} establishes a
connection of the Bethe free energy approach to the loop calculus
and LP decoding. Section \ref{sec:test} unveils an underlying loop
structure,  e.g. emergence of certain critical loops, for the family
of instantons that appear in the $(155,64,20)$ code performing over
the AWGN channel and decoded by LP \cite{06CS}. The effective free
energy approach, suggesting modification of the BP gauges to account
for the critical loop, is introduced in Section \ref{sec:Eff}. An
improved LP decoding, called LP-erasure, is presented in Section
\ref{sec:LP-er}. Numerical test of the LP-erasure algorithm using an
example of the $(155,64,20)$ code is discussed in Subsection
\ref{subsec:test2} where we also demonstrate that all previously
found bare LP instantons (most damaging noise configurations) are
actually corrected by the LP-erasure procedure thus completely
reducing the error-floor observed for the standard LP decoding. The
final Section \ref{sec:Con} contains conclusions and discussions.

\section{Introduction}
\label{sec:Intro}

\subsection{Decoding in terms of Statistical Inference}
\label{subsec:SI}

A message word that consists of $K$ bits is encoded to an $N$-bit
long codeword, $N>K$. In the binary linear case the code can be
conveniently represented by $M\geq N-K$ constraints, usually
referred to as parity checks or simply checks. Formally, ${\bm\pi} =
(\pi_{1}, \dots, \pi_{N})$ with $\pi_{i} = \pm 1$, is one of the
$2^{K}$ codewords if and only if $\prod_{i\in \alpha}\pi_{i} = 1$
for all checks $\alpha = 1, \dots, M$, where $i\in\alpha$ if the bit
$i$ contributes the check $\alpha$. The relation between bits and
checks (we use $i\in\alpha$ and $\alpha\ni i$ interchangeably) is
often described in terms of an $M\times N$ parity-check matrix
$\hat{ H}$ that consists of ones and zeros: $H_{\alpha i}=1$ if
$i\in\alpha$ and $H_{\alpha i}=0$ otherwise. A bipartite graph
representation of $\hat{H}$, with bits marked as circles, checks
marked as squares and edges corresponding to the corresponding
nonzero elements of $\hat{H}$, is usually called the Tanner graph
associated with the code. For an LDPC code $\hat{H}$ is sparse, i.e.
most of the entries are zeros. Transmitted through a noisy channel,
a codeword gets corrupted due to the channel noise, so that the
channel output at the receiver is ${\bm x}\neq{\bm\pi}$. Even though
an information about the original codeword is lost at the receiver,
one still possesses the full probabilistic information about the
channel, i.e. the conditional probability $P({\bm x}|{\bm \sigma})$
for a codeword ${\bm \sigma}$ to be a pre-image for the output word
${\bm x}$ is known. In the case of independent noise samples the
full conditional probability can be decomposed into a product,
$P({\bm x}|{\bm \sigma})=\prod_i p(x_i|\sigma_i)$. The channel
output at a bit can be conveniently characterized by the so-called
log-likelihood $h_i=\log(p(x_i|+1)/p(x_i|-1))/2s^2$ measured in the
units of the Signal-to-Noise-Ratio (SNR), normally defined as
$2s^2$. For a common model of the Additive White Gaussian Noise
(AWGN) channel , $p(x|\sigma) = \exp(-s^2 (x -
\sigma)^2/2)/\sqrt{2\pi/s^2}$, $h_i=x_i$. The decoding goal is to
infer the original message from the received output ${\bm x}$. ML
decoding (that generally requires an exponentially large number
$2^K$ of steps) corresponds to finding ${\bm \sigma}$ maximizing the
following weight (probability distribution) function
\begin{eqnarray}
W({\bm \sigma})=Z^{-1}\prod\limits_\alpha
\delta\left(\prod\limits_{i\in\alpha}\sigma_i,+1\right)
 \exp\left(\sum\limits_i\sigma_i h_i\right),
\label{WLDPC}
\end{eqnarray}
where the normalization factor $Z$ that  enforces the
$\sum_{\bm\sigma} W({\bm\sigma})=1$ condition is called the
partition function in the statistical physics literature.
Maximum-A-Posteriori (MAP) decoding boils down to finding
a-posteriori log-likelihood (magnetization) at a bit defined
according to
\begin{eqnarray}
{\bm m}=\sum\limits_{\bm\sigma}{\bm \sigma}W({\bm\sigma}),
 \label{magLDPC}
\end{eqnarray}
followed by taking an absolute value of the result bit-wise.

Even though bits and checks play an essentially different role in
the LDPC decode, it actually turns out to be formally convenient to
consider them on equal footing thus putting the relevant inference
problem in a more general context of graphical models
\cite{01KFL,01For,01Loe} where binary variables are shifted from
bits/vertexes to edges of the corresponding Tanner graph.

A general vertex model, also called normal factor graph model
\cite{01For}, is determined by the weight (probability distribution)
function ${\cal W}$, which along with the partition function $Z$ can
be represented in the following general form
\begin{eqnarray}
\label{VM-weight-gen}
 W({\bm\sigma})= Z^{-1}\prod\limits_{a\in X}f_a({\bm\sigma}_a),\ \
 Z=\sum\limits_{\bm \sigma} \prod\limits_{a\in
 X}f_a({\bm\sigma}_a),
\end{eqnarray}
where $X$ defines a graph consisting of vertexes and edges; $a$
denotes a node (vertex) in the model; an elementary spin resides at
the edge connecting two neighboring vertexes, $\sigma_{ab}=\pm 1$,
for $b\in a$ and $a\in b$; ${\bm \sigma}_a$ stands for the vector
built of all $\sigma_{ab}$ with $b\in a$; ${\bm \sigma}$ is a
particular configuration of spins on all the edges. With this
notation one assumes that $\sigma_{ab}=\sigma_{ba}$.

The problem of LDPC decoding (\ref{WLDPC}) represents a particular
case of the general vertex model with bits and checks combined in
one family of vertexes $\{a\}=\{i\}\cup\{\alpha\}$, the Tanner graph
$X$ and the factor functions defined according to
\begin{eqnarray}
 &&\!\!\!\!\!\! f_i({\bm \sigma}_i)=\left\{
 \begin{array}{cc}\!\! \exp(h_i\sigma_i), & \!\!\sigma_{i\alpha}=\sigma_{i\beta}=\sigma_i
 \ \ \forall \alpha,\beta\ni i\ \ \\
 0, & \mbox{  otherwise;}\end{array}\right. \label{fi_LDPC}\\
 && \!\!\!\!\!\! f_\alpha({\bm \sigma}_\alpha)=\left\{
 \begin{array}{cc} 1, & \prod_{i\in\alpha}\sigma_i=1\ \
 \\ 0, & \prod_{i\in\alpha}\sigma_i=-1\end{array}\right.. \label{fa_LDPC}
\end{eqnarray}

Our strategy will be to keep the general vertex model notations
whenever possible. The transition in the general formulas to our
focus case of the LDPC decoding  will always be simple and
straightforward according to Eqs.~(\ref{fi_LDPC},\ref{fa_LDPC}).

\subsection{Loop Calculus \cite{06CCa,06CCb}}
\label{subsec:LC}

Consider a  general vertex model (\ref{VM-weight-gen}) and relax the
condition $\sigma_{ab}=\sigma_{ba}$, thus treating $\sigma_{ab}$ and
$\sigma_{ba}$ as independent binary variables. We represent the
partition function in the form:
\begin{eqnarray}
\label{Z-gen-graph}
Z=\sum_{\bm\sigma}\prod_{a}f_{a}({\bm\sigma}_{a})\prod_{bc}\frac{1+\sigma_{bc}\sigma_{cb}}{2}.
\end{eqnarray}
Note that for this representations the vectors ${\bm\sigma}_{a}$
become independent variables. Also in the product over edges,
$(bc)$, we assume that each edge contributes only once. We further
introduce a parameter vector ${\bm\eta}$ with the set of independent
components $\eta_{ab}$. Making use of the algebraic relation
\begin{eqnarray}
 && \frac{\cosh(\eta+\chi)(1+\pi\sigma)}{(\cosh\eta+\sigma\sinh\eta)(\cosh\chi+\pi\sinh\chi)}
 =1+\label{edge-relation}\\ &&
 \left(\tanh(\eta+\chi)-\sigma\right)\left(\tanh(\eta+\chi)-\pi\right)\cosh^{2}(\eta+\chi).
 \nonumber
\end{eqnarray}
we arrive at the following representation for the partition function
that is ready for a subsequent loop decomposition
\begin{eqnarray}
 &&
 Z=\left(\!\prod_{bc}2\cosh\left(\eta_{bc}+\eta_{cb}\right)\!\right)^{-1}\!\!
 \sum_{\bm\sigma}\prod_{a}P_{a}
 \prod_{bc}V_{bc},\label{Z-ready-gen}\\ &&
 P_{a}({\bm\sigma}_{a})=f_{a}({\bm\sigma}_{a})\prod_{b\in a}
 \left(\cosh\eta_{ab}+\sigma_{ab}\sinh\eta_{ab}\right); \;\;\;
 \label{Gen_Pj} \\ &&
 V_{bc}\left(\sigma_{bc},\sigma_{cb}\right)=1+\left(\tanh(\eta_{bc}+\eta_{cb})-\sigma_{bc}\right)
 \nonumber\\ && \ast\left(\tanh(\eta_{bc}+\eta_{cb})-\sigma_{cb}\right)\cosh^{2}(\eta_{bc}+\eta_{cb}).
\label{Vbc}
\end{eqnarray}

Eqs.~(\ref{Z-ready-gen},\ref{Gen_Pj},\ref{Vbc}) are generally valid
for any choice of the $\eta$ fields (gauge choice). However, in the
rest of this paragraph we will be discussing a particular choice of
the gauge fields, $\eta^{({\it bp})}$, special because of its
relation to Belief Propagation. The desired decomposition is
obtained by expanding the $V$-terms followed by a local computation.
The parameters (gauges) ${\bm\eta}$ are chosen using the criterion
that subgraphs with at lest one loose end do not contribute to the
decomposition. This can be achieved if the parameters satisfy the
following system of equations:
\begin{eqnarray}
\label{General-BP-gen}
\sum_{{\bm\sigma}_{a}}\left(\tanh(\eta_{ab}^{({\it
bp})}+\eta_{ba}^{(\it
bp)})-\sigma_{ab}\right)P_{a}({\bm\sigma}_{a})=0.
\end{eqnarray}

It is also straightforward to check that the gauge-fixing condition
Eq.(\ref{General-BP-gen}) corresponds to an extremum of $Z_0$,
\begin{equation}
\Biggl.\frac{\delta Z_0}{\delta\eta_{ab}}\Biggr|_{\eta^{({\it
bp})}}=0, \label{extrZ0}
\end{equation}
where
\begin{eqnarray}
Z_0=\left.\left(\prod_{bc}2\cosh\left(\eta_{bc}+\eta_{cb}\right)\right)^{-1}
\sum_{\bm\sigma}\prod_{a}P_{a}({\bm\sigma}_{a})\right|_{\eta^{(bp)}},
\label{Z0}
\end{eqnarray}
is the bare part of the partition function $Z$ derived from
Eq.~(\ref{Z-ready-gen}) with the product of the vertex $V$-terms
replaced by unity.

\begin{figure} [b]
\includegraphics[width=0.45\textwidth]{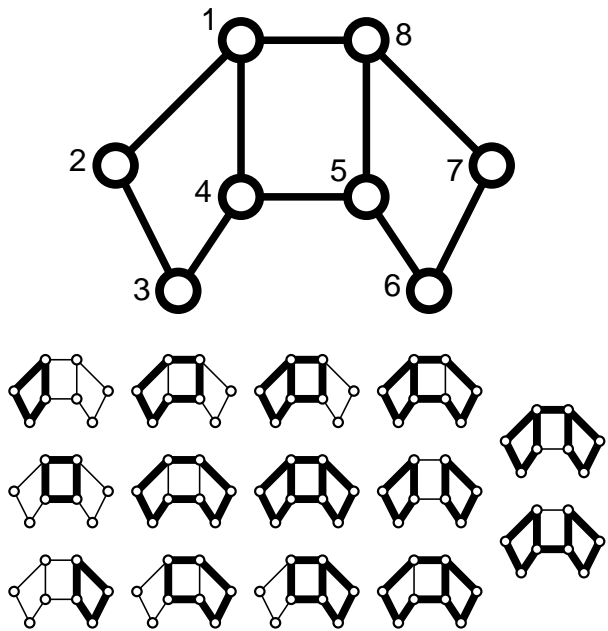}
\caption{Example of a general vertex model. Fourteen possible marked paths (generalized loops) for
the example are shown in bold at the bottom.} \label{fig:paths}
\end{figure}

Eqs.~(\ref{General-BP-gen}) constitute the BP system of equations,
represented in terms of parameters $\eta^{(bp)}$. Calculated within
BP the probability of finding the whole family of edges connected to
a node $a$ in the state ${\bm \sigma}_a$ is
\begin{eqnarray}
 b_a^{({\it bp})}({\bm \sigma}_a)=\left.\frac{P_{a}({\bm\sigma}_{a})}{\sum_{{\bm
 \sigma}_a}
 P_a({\bm\sigma}_a)}\right|_{\eta^{(bp)}}.
\label{gen_prob_BP}
\end{eqnarray}
A typical sum, entering a diagram contribution for a generalized
loop ${\it C}$, is expressed in terms of the corresponding
irreducible correlation functions of the spin variables computed
within BP:
\begin{eqnarray}
 \mu_{a}^{({\it bp})}= \sum_{{\bm\sigma}_{a}}b_a^{({\it bp})}({\bm\sigma}_{a})
 \prod\limits_{b\in a,{\it C}}
 \left(\sigma_{ab}-m_{a b}^{({\it bp})}\right),
 \label{gen-mu}
\end{eqnarray}
where $m_{ab}^{({\it bp})}$ is the magnetization (a-posteriori
log-likelihood) at the edge $(ab)$ calculated within BP
\begin{eqnarray}
 m_{ab}^{({\it bp})}=\sum_{{\bm\sigma}_{a}} b_a^{({\it bp})}({\bm\sigma}_{a}) \sigma_{ab}.
 \label{gen-m}
\end{eqnarray}
Making use of Eqs.~(\ref{Z-ready-gen},\ref{gen-mu},\ref{gen-m}) one
derives the following final expression for the partition function
\begin{eqnarray}
 && Z=Z_0\left(1+\sum\limits_{\it C} r({\it C})\right),\nonumber\\
 &&
 r({\it C})=\frac{\prod\limits_{a\in{\it C}}\mu_a^{({\it bp})}}
{\prod\limits_{(ab)\in C}(1-(m_{ab}^{({\it bp})})^2)},
\label{gv_ser}
\end{eqnarray}
where $Z_0$ is taken at $\eta^{({\it bp})}$ and the summation in Eq.~(\ref{gv_ser}) runs over all
allowed ${\it C}$ (marked) paths in the graph associated with the model; $(ab)$ marks the edge on
the graph that connects nodes $a$ and $b$. A marked path is allowed to branch at any node/vertex,
however it cannot terminate at a node. We refer to such a structure as a loop (it is actually some
kind of a generalized loop since branching is allowed; we use the shorter name for convenience). An
example is given in Fig.~(\ref{fig:paths})

In the LDPC case (\ref{fi_LDPC},\ref{fa_LDPC}) the loop series expressions (\ref{gv_ser}) assumes
the following form
\begin{eqnarray} &&
 Z_{\it\tiny LDPC}=Z_0\left(1+\sum\limits_{{\it
C}}r({\it C}) \right),\label{Zseries}\\ &&
 r({\it C})=\prod\limits_{i,\alpha\in {\it C}}\mu_{\alpha}^{(bp)}\mu^{(bp)}_{i},
 \quad q_i=\sum\limits_{\alpha\in{\it C}}^{\alpha\ni i}1,
\nonumber\\
 && \mu^{(bp)}_{i}=\frac{(1-m^{(bp)}_i)^{q_i-1}+(-1)^{q_i}(1+m^{(bp)}_i)^{q_i-1}}
 {2(1-(m_i^{(bp)})^2)^{q_i-1}},
\nonumber\\
&& \hspace{-0.7cm}
 \mu_{\alpha}^{(bp)}\!=\!\sum\limits_{\sigma_\alpha}\!b_\alpha^{(bp)}(\sigma_\alpha)
 \!\!\prod\limits_{i\in{\it C}}^{i\in\alpha}\!(\sigma_i\!-\!m_i^{(bp)}),\;
m_i^{(bp)}\!=\!\sum\limits_{\sigma_i}\!
b_i^{(bp)}(\sigma_i)\sigma_i, \nonumber
\end{eqnarray}
where $b^{(bp)}_i(\sigma_i)$ and $b_\alpha^{(bp)}({\bm\sigma}_\alpha)$ are the beliefs defined on
bits and checks respectively according to
\begin{eqnarray}
&& b_i^{(bp)}(\sigma_i)\propto
\exp\left(\frac{\sigma_i}{q_i-1}\left(\sum\limits_{\alpha\ni
i}\eta^{(bp)}_{i\alpha}-h_i\right)\right),\label{bi}\\ &&
b_\alpha^{(bp)}({\bm\sigma}_\alpha)\propto\delta\left(\prod\limits_{i\in\alpha}\sigma_i,+1\right)
\exp\left(\sum\limits_{i\in\alpha}\eta_{i\alpha}\sigma_i\right),
\label{balpha}\\ && \eta_{i\alpha}^{(bp)}=h_i+\sum\limits_{\beta\ni
i}^{\beta\neq \alpha}
 \tanh^{-1}\left(\prod\limits_{j\in\beta}^{j\neq
 i}\tanh\eta_{j\beta}^{(bp)}\right).\label{BPia}
\end{eqnarray}
Eq.~(\ref{BPia}) represents a traditional form of the BP equations for LDPC codes
\cite{63Gal,68Gal}.

See \cite{06CCb} for more details of the derivations sketched in
this Subsection.

\subsection{Calculating a-posteriori log-likelihoods}
\label{subsec:mag}

Formally a-posteriori log-likelihood (magnetization) is defined by
\begin{eqnarray}
 m_{ab}\equiv \sum\limits_{\bm \sigma}\sigma_{ab}W({\bm\sigma}). \label{mag}
\end{eqnarray}
Full magnetization at an edge of a general graphical model can be
recalculated using Eqs.~(\ref{gv_ser},\ref{mag}) with the gauge
fields fixed according to the BP rules (\ref{General-BP-gen}). There
are two complementary ways to perform these calculations. One can
incorporate infinitesimal test-field, $h_{ab}$, into the factor
functions by $f_a\to f_a\exp(\sigma_{ab} h_{ab})$, and use the
generalized partition function to generate the corresponding
magnetizations by simply differentiating $\ln Z$ (with $Z$ taken in
the loop-series representation) with respect to the test-field
followed by taking the $h_{ab}\to 0$ limit. Alternatively and
straightforwardly, one can calculate the magnetization according to
the definition (\ref{mag}) with the probability measure taken
according to Eq.~(\ref{Z-ready-gen}) in the following BP-gauge form
\begin{eqnarray}
W^{(bp)}({\bm\sigma})\!=\!\left(\prod_{a}P_{a}
 \prod_{bc}\frac{V_{bc}}{2\cosh\left(\eta_{bc}\!+\!\eta_{cb}\right)}
 \right)_{\eta^{(bp)}}.\label{W-BP}
\end{eqnarray}
Applying the same "marked path" rule to calculate the magnetization
as we used in the previous Subsection to derive the loop-series
expression for the partition function, we arrive at
\begin{eqnarray}
 && \hspace{-1.2cm} m_{ab}=\frac{m_{ab}^{({\it bp})}\left(1+
 \sum\limits_{{\it C}}^{a\notin{\it C}} r({\it C})\right)
 +\sum\limits_{{\it C}_{a\to b}}\delta m^{({\it bp})}_{a\to b;{\it C}_{a\to b}}}
 {1+\sum\limits_{{\it C}} r({\it C})},\label{m_full}\\
 && \hspace{-1.2cm} \delta m_{a\to b;{\it C}}^{({\it bp})}=\frac{\mu_{a\to b;{\it C}}^{({\it bp})}
 \prod\limits_{c\in{\it C}}^{c\neq a}\mu_c^{({\it bp})}}
 {\prod\limits_{(ab)\in {\it C}}(1-(m_{ab}^{({\it
 bp})})^2)},\label{m1}\\ &&
 \hspace{-1.2cm} \mu_{a\to b;{\it C}}^{({\it bp})}=\frac{
 \sum\limits_{{\bm \sigma}_a}b_a^{({\it bp})}({\bm\sigma}_a)\sigma_{ab}
 \prod\limits_{c\in a,C}(\sigma_{ac}-m_{ac}^{({\it bp})})}{
 1-(m_{ab}^{({\it bp})})^2},
 \label{mua;ab}
\end{eqnarray}
where ${\it C}_{a\to b}$ consists of an extended family of loops with the connectivity degree at
node $a$ being one or higher while the connectivity degree at node $b$, as well as at any other
node that belongs to ${\it C}_{a\to b}$ and different from $a$, being two or higher.

Note that for any generalized loop ${\it C}$ there may be many extended loops ${\it C}_{ab}$
propagating correlations imposed by the local loop contribution all over the graph.

Replacing the hole families of the generalized loops ${\it C}$ and the extended generalized loops
${\it C}_{a\to b}$ in Eq.~(\ref{m1}) by some relevant subfamilies  constitutes an approximation
which provides an improvement over the bare BP approximation.

\subsection{Bethe-Free energy and Linear Programming Decoding}
\label{subsec:BFE}

The variational approximation for a general vertex model reads as
follows \cite{05YFW}. The Bethe free energy
\begin{eqnarray}
 F=\sum_a \sum_{{\bm \sigma}_a}b_a
 \ln\left(\frac{b_a}{f_a}\right)-\sum_{(a b)}
 \sum_{\sigma_{ab}}
 b_{ab}
 \ln b_{ab},
 \label{Bethe}
\end{eqnarray}
is a functional of the beliefs $b_a({\bm \sigma}_a), b_{ac}(\sigma_{ac})$, defined on the vertices
and edges of the graph respectively. BP equations can be introduced as equations for a conditional
extremum of the Bethe-free energy.  The realizability conditions (constraints) are
\begin{eqnarray}
 && \!\!\!\!\!\forall\ a,c;\ c\in a:\ \ 0\leq b_a({\bm \sigma}_a),b_{ac}(\sigma_{ac})\leq 1,\label{ineq}\\
 && \!\!\!\!\!\forall\ a,c;\ c\in a:\ \ \sum_{{\bm \sigma}_a} b_a({\bm \sigma}_a)=
 \sum_{\sigma_{a,c}} b_{ac}(\sigma_{ac})=
 1,\label{norm}\\
 && \quad\quad\quad b_{ac}(\sigma_{ac})\!\!=\!\!\!
 \sum_{{\bm \sigma}_a\setminus\sigma_{ac}}\!\!\!\! b_a({\bm
 \sigma}_a)\!=\!\!\!
 \sum_{\sigma_c\setminus\sigma_{ac}}\!\!\! b_c({\bm \sigma}_c),\label{cons}
\end{eqnarray}
where we assume $\sigma_{ac}=\sigma_{ca}$. The second term on the rhs of Eq.~(\ref{Bethe}) is the
entropy term that corrects for ``double counting" of the link contribution: any link appears twice
in the entropy part of the first term on the rhs of Eq.~(\ref{Bethe}).

Optimal configurations of beliefs minimize the Bethe free energy
(\ref{Bethe}) subject to the constraints
(\ref{ineq},\ref{norm},\ref{cons}). Introducing the constraints as
Lagrange multipliers in the effective Lagrangian and looking for the
extremum with respect to all possible beliefs leads to
\begin{eqnarray}
 && b_a^{(bp)}({\bm\sigma}_a)\propto f_a({\bm \sigma}_a)\prod\limits_{b\in
 a}\exp(\eta_{ab}^{(bp)}\sigma_{ab}),\label{ba}\\
&& b_{ab}^{(bp)}(\sigma_{ab})\propto
\exp((\eta_{ab}^{(bp)}+\eta_{ba}^{(bp)})\sigma_{ab}), \label{bab}
\end{eqnarray}
where $\propto$ indicates that one should use the normalization conditions (\ref{norm}) to
guarantee that the beliefs sum up to one. It is straightforward to check that the system of
Eqs.~(\ref{ba},\ref{bab}) supplemented with the normalization and consistency conditions for the
beliefs Eqs.~(\ref{norm},\ref{cons}) is fully equivalent to the BP equations (\ref{General-BP-gen})
discussed above.

If the aforementioned optimization procedure is performed with the compatibility conditions
(\ref{cons}) excluded, yet all other constraints accounted for, one still obtains
Eqs.~(\ref{ba},\ref{bab}) for the beliefs with $\eta^{(bp)}$ replaced by yet unconditioned $\eta$.
Expressing the beliefs in terms of the $\eta$-fields according to the relaxed version of
Eqs.(\ref{ba},\ref{bab}) we arrive at the following expression for the Bethe free energy in terms
of the $\eta$ variables:
\begin{eqnarray}
 && \!\!\!\!\!\!\!\!F\!=\!{\cal F}_0
 \!+\!\sum\limits_{(ab)}\left(\eta_{ab}m_{a\to b}^{(*)}\!+\!
 \eta_{ba}m_{b\to a}^{(*)}\!-\!(\eta_{ab}+\eta_{ba})m_{ab}^{(*)}\right),\nonumber\\
 &&
 m_{a\to b}^{(*)}\equiv
 \frac{\sum\limits_{{\bm \sigma}_a}\sigma_{ab}f_a({\bm \sigma}_a)\prod\limits_{c\in
 a}\exp(\eta_{ac}\sigma_{ac})}
 {\sum\limits_{{\bm \sigma}_a}f_a({\bm \sigma}_a)\prod\limits_{c\in
 a}\exp(\eta_{ac}\sigma_{ac})}
 ,\label{m1ab}\\
 && m_{ab}^{(*)}\equiv
 \tanh\left(\eta_{ab}+\eta_{ba}\right),
 \label{m2ab}
\end{eqnarray}
where ${\cal F}_0(\eta)\equiv-\ln Z_0(\eta)$ and $m_{a\to b}^{(*)}$ and $m_{ab}^{(*)}$ are two
expressions for a-posteriori log-likelihoods (magnetizations) at the edge $(ab)$ which are equal to
each other if the compatibility conditions are accounted for. For an arbitrary choice of $\eta$ the
magnetizations are different. However, for the special choice of $\eta$ that corresponds to
$\eta^{({\it bp})}$ the two generally different magnetizations become equal. More formally: the
Belief Propagation equations can be expressed as
\begin{eqnarray}
m_{a\to b}^{(*)}|_{(*)\to ({\it bp})}=m_{b\to a}^{(*)}|_{(*)\to
({\it bp})}
 =m_{ab}^{(*)}|_{(*)\to ({\it bp})}, \label{BP_mag}
\end{eqnarray}
satisfied $\forall ab$, where $\eta$ and $b^{(*)}$ turn into $\eta^{({\it bp})}$ and $b^{({\it
bp})}$, respectively.

There is also a relation between the Bethe free energy minimization
approach and the LP decoding \cite{03FWK,03KV,04VK,05VK}. Consider
the Bethe free energy Eq.~(\ref{Bethe}) in the $f\to\infty$ limit of
zero noise or, equivalently, large SNR. In this limit the
self-energy contribution to the free energy dominates the entropy
terms and the latter can be safely neglected. However, both the self
energy and the constrains (\ref{ineq},\ref{norm},\ref{cons}) are
linear in beliefs. Therefore this asymptotic optimization problem
can be solved efficiently by means of the standard linear
programming approach.

\section{Loop calculus analysis of the LP-instantons for the
$(155,64,20)$ code}
 \label{sec:test}

We consider a family of ($\sim 200$) instantons, i.e. noise
configurations (log-likelihoods) corresponding to the
pseudo-codewords with the effective distance smaller then the
Hamming distance of the code.  This set of instantons and related
pseudo-codewords was found in \cite{06CS} for the Tanner
$(155,64,20)$ code performing over the Additive-White-Gaussian-Noise
(AWGN) channel and decoded using LP decoding.

In short the method/algorithm of \cite{06CS}, called pseudo-codeword
search algorithm, proposes an efficient way of describing LP
decoding polytope and the pseudo-codeword spectra of the code. It
approximates a pseudo-codeword and the corresponding noise
configuration on the error-surface surrounding the zero codeword
correspondent to the shortest effective distance of the code. The
algorithm starts from choosing a random initial noise configuration.
The configuration is modified through a discrete number of steps.
Each step consists of two sub-steps. First, one applies LP decoder
to the initial noise-configuration deriving a pseudo-codeword.
Second, one finds the noise configuration equidistant from both the
pseudo codeword and the zero codeword. The resulting noise
configuration is used as an entry for the next step. The algorithm,
tested on the Tanner $(155,64,20)$ code and Margulis $p=7$ and
$p=11$ codes ($672$ and $2640$ bits long respectively) shows very
fast convergence.

Discussing the instantons, found for the $(155,64,20)$ code with the
pseudo-codeword-search algorithm, one after another we aim to
associate for each instanton the corresponding critical loop,
$\Gamma$ that generates a contribution to the loop series
(\ref{Zseries}), comparable to the bare BP contribution.

We restrict our search for the critical loop contribution to the class of single-connected loops,
i.e. loops that consist of checks and bits with each check connected to only two bits of the loop.
According to Eqs.~(\ref{Zseries}) such contribution to the loop series is the product of all the
triads, $\tilde{\mu}^{(bp)}$, along the loop,
\begin{eqnarray}
 && r(\Gamma)\!=\!\prod_{\alpha\in\Gamma}\tilde{\mu}_\alpha^{(bp)},\label{rGamma}\\
 && \tilde{\mu}_{\alpha}^{(bp)}\!=\!\frac{\mu_{\alpha}^{(bp)}}
 {\sqrt{(1\!-\!(m_i^{(bp)})^2)(1\!-\!(m_j^{(bp)})^2)}},
\label{tildemu}
\end{eqnarray}
where for any check $\alpha$ that belongs to $\Gamma$, $i,j$ is the
only pair of $\alpha$ bit neighbors that also belongs to $\Gamma$.
By construction, $|\tilde{\mu}_{\alpha;ij}^{(bp)}|\leq 1$. We
immediately find that for the critical loop contribution to be
exactly equal to unity (where unity corresponds to the bare BP
term), the critical loop should consist of triads with all
$\tilde{\mu}^{(bp)}$ equal to unity by absolute value. Even if
degeneracy is not exact one still anticipates the contributions from
all the triads along the critical loop to be reasonably large, as an
emergence of a single triad with small $\tilde{\mu}^{(bp)}$ will
make the entire product negligible in comparison with the bare BP
term. This consideration suggests that an efficient way to find a
single connected critical loop, $\Gamma$, with large  $|r(\Gamma)|$
consists of, first, ignoring all the triads with
$|\tilde{\mu}^{(bp)}|$ below a certain $O(1)$ threshold,  say
$0.999$, and second checking if one can construct a single connected
loop out of the remaining triads. If no critical loop is found we
lower the threshold till a leading critical loop emerges.

Applied to the set of instantons of the Tanner $(155,64,20)$ code
with the lowest effective distances this triad-based scheme
generates $r(\Gamma)$ that is exactly unity by the absolute value.
This is the special degenerate case when the critical loop
contribution and the BP/LP contribution are equal to each other by
absolute value. Thus only the sixth of the first dozen of instantons
has $r(\Gamma)\approx 0.82$ while all others yield $r(\Gamma)=1$. To
extend the triad-based search scheme to the instantons with larger
effective distance one needs to decrease the threshold. This always
results in emergence of at least one single connected loop with
$r(\Gamma)\sim 1$. Note, that it may be advantageous, even thought
not necessary, to include in this triad-based search for the
critical loop some additional criteria. In particular, we found it
useful to also require that the absolute values of the a-posteriori
log-likelihoods of the bits involved in the critical loop be larger
than certain threshold.

\begin{figure}

\subfigure{
\includegraphics[width=0.45\textwidth]{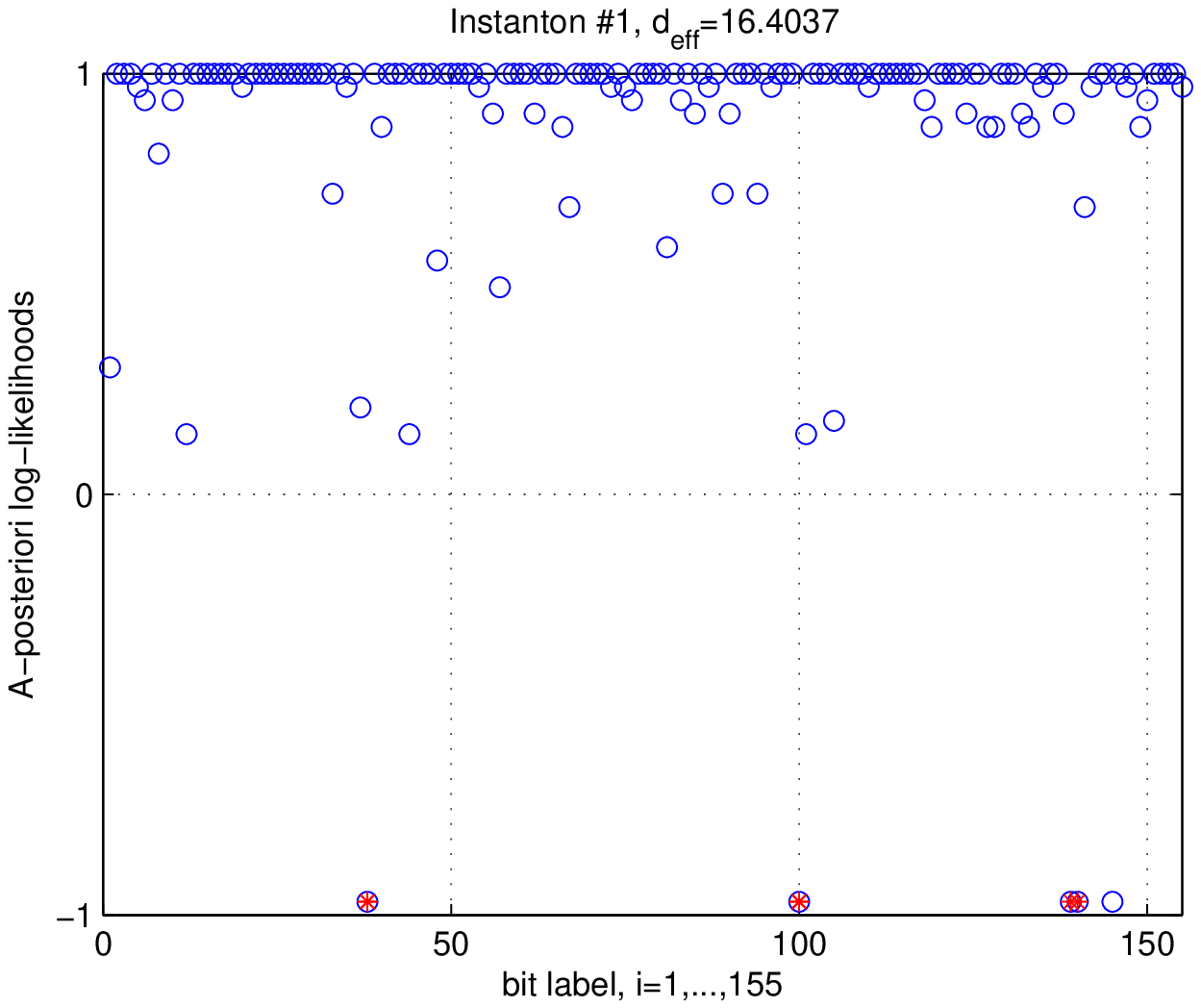}
\includegraphics{N1s.eps}

\includegraphics[width=0.45\textwidth]{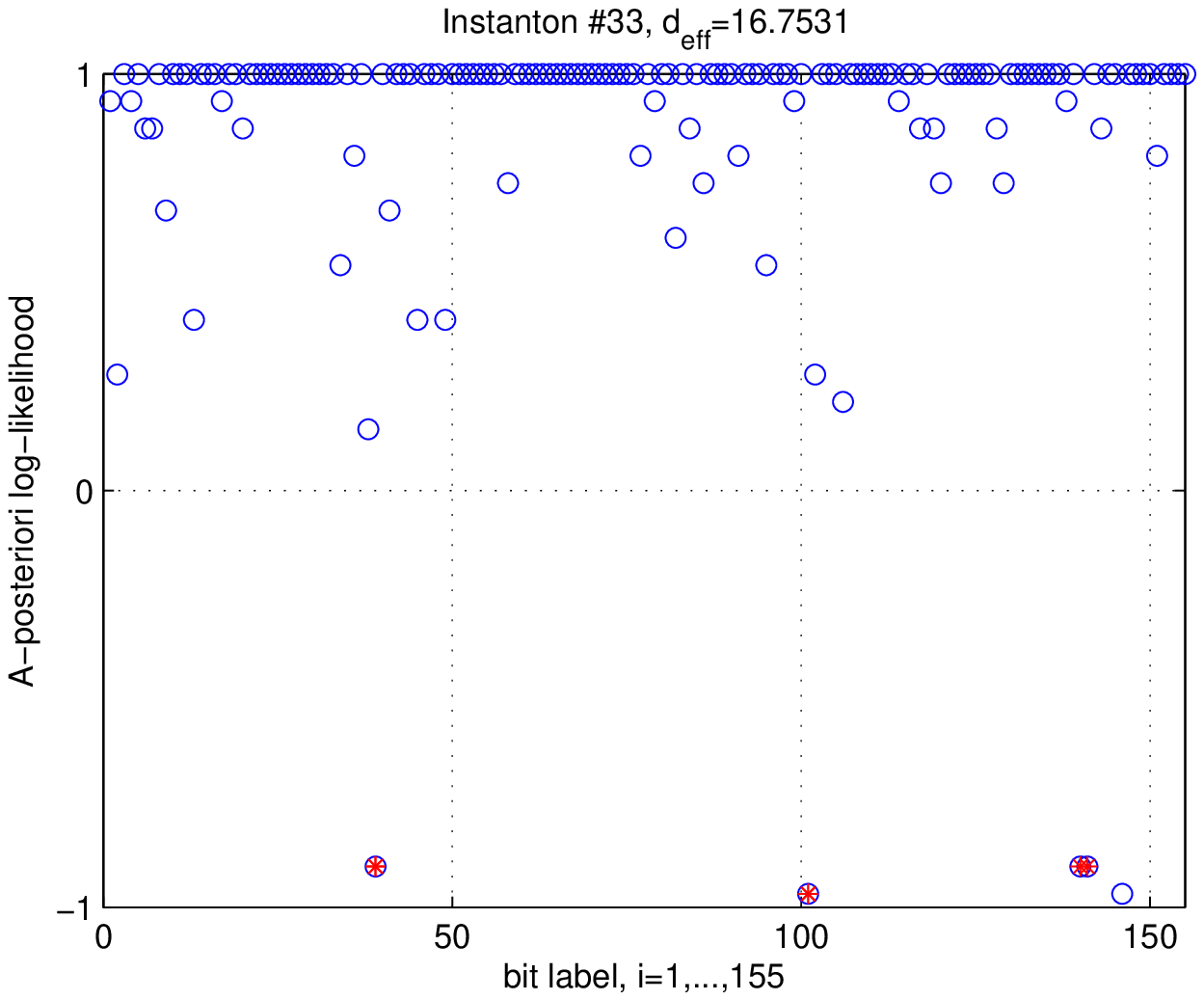}
\includegraphics{N33s.eps}}

\subfigure{
\includegraphics[width=0.45\textwidth]{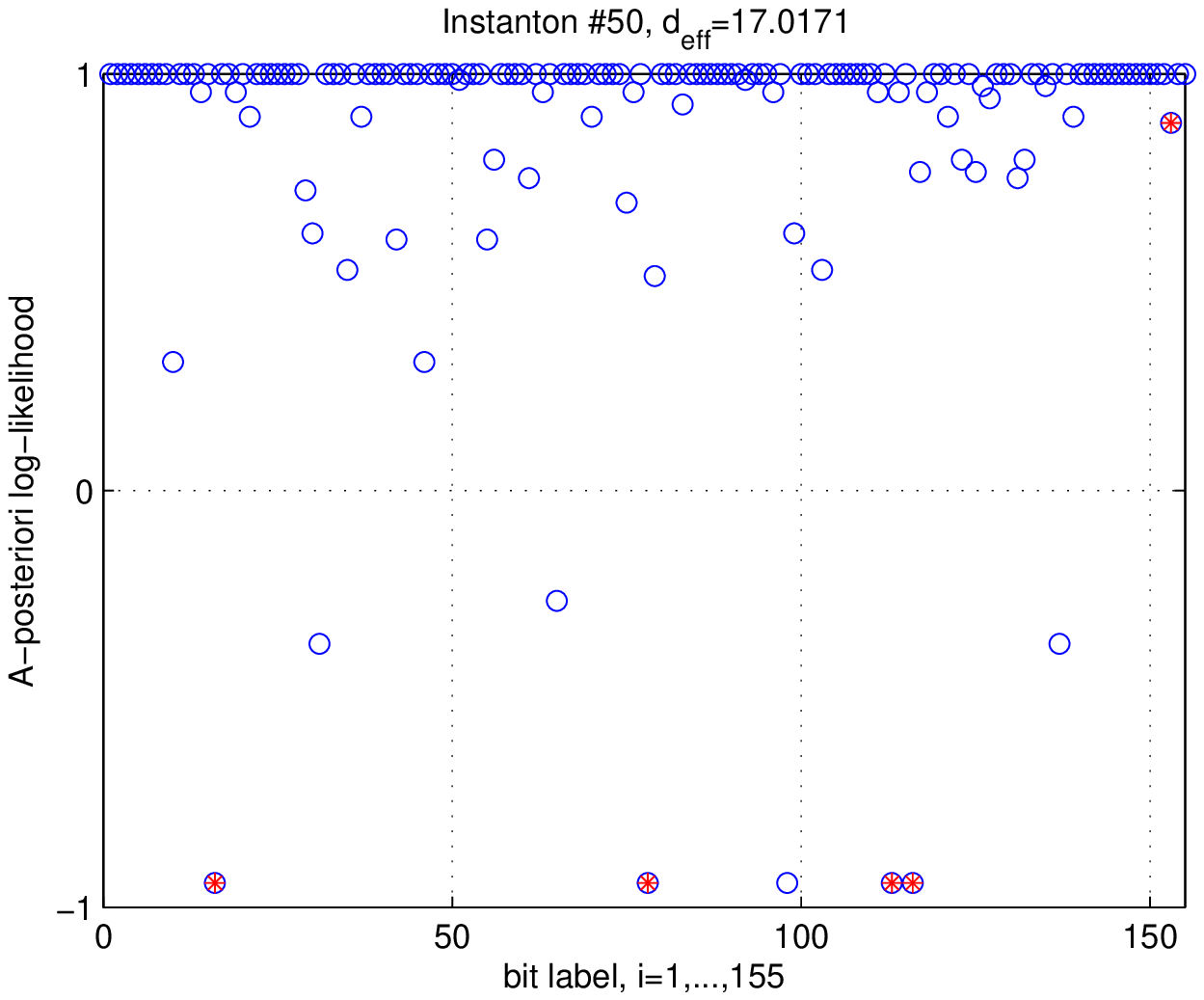}
\includegraphics{N50s.eps}

\includegraphics[width=0.45\textwidth]{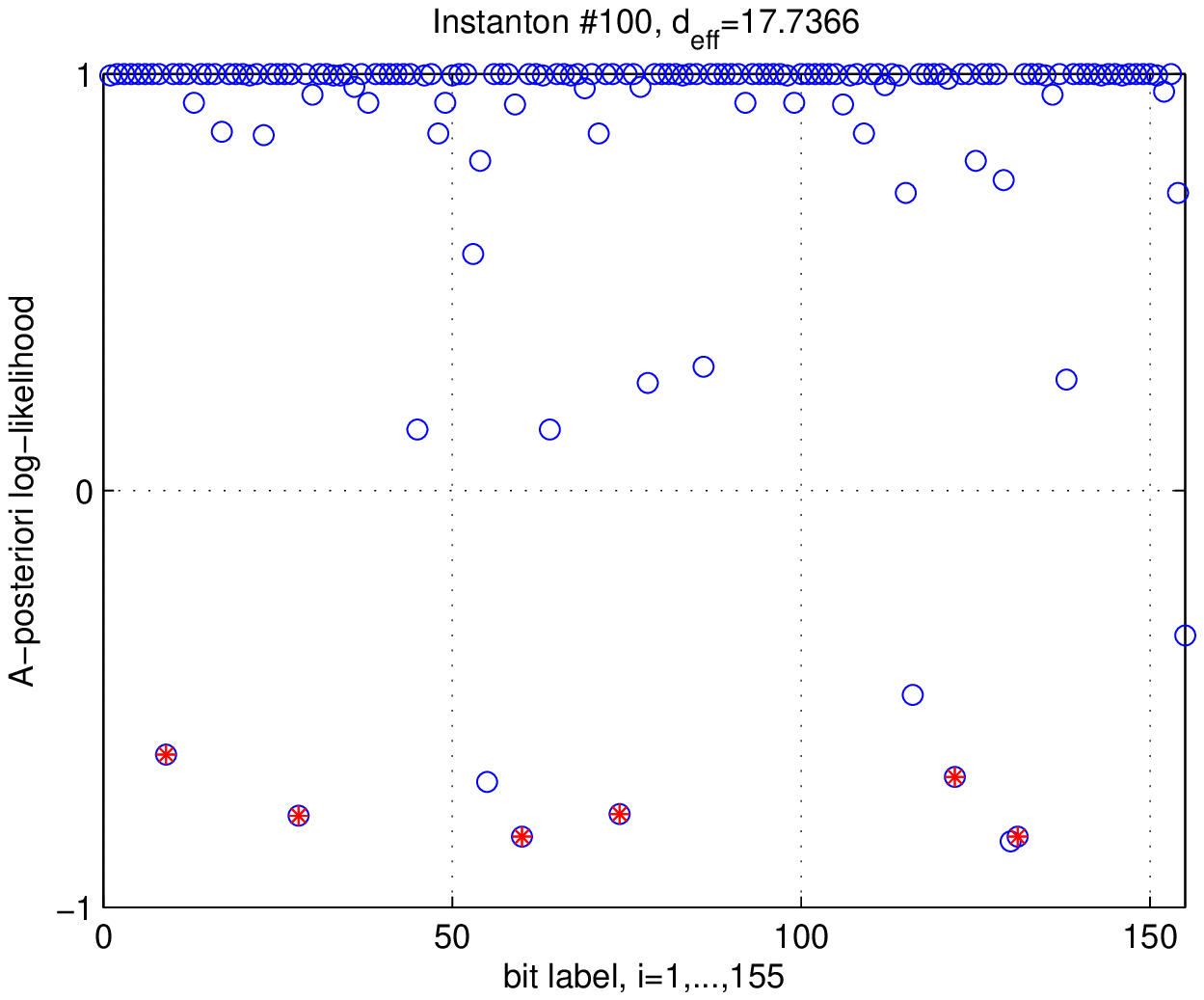}
\includegraphics{N100s.eps}}

\subfigure{
\includegraphics[width=0.45\textwidth]{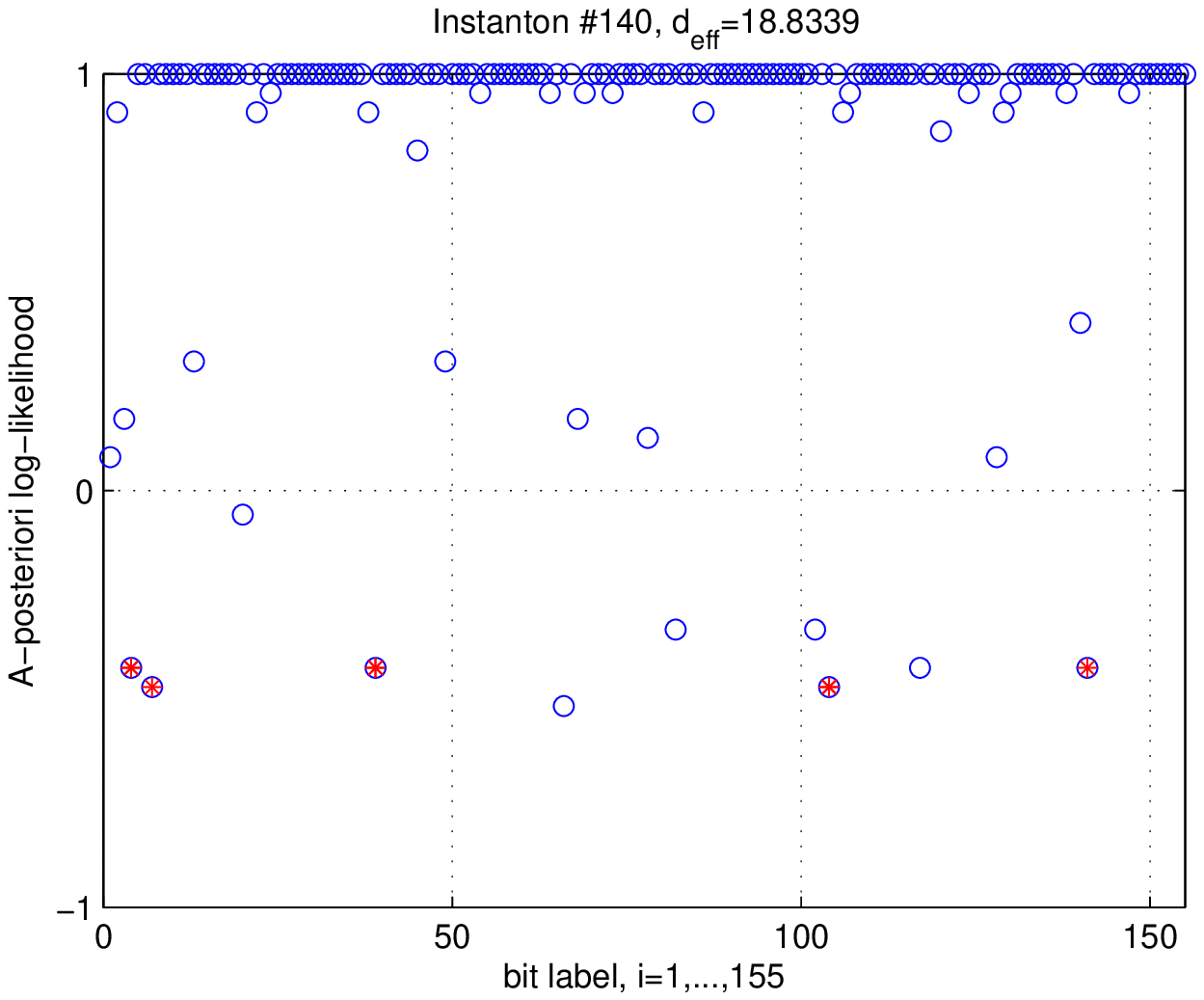}
\includegraphics{N140s.eps}

\includegraphics[width=0.45\textwidth]{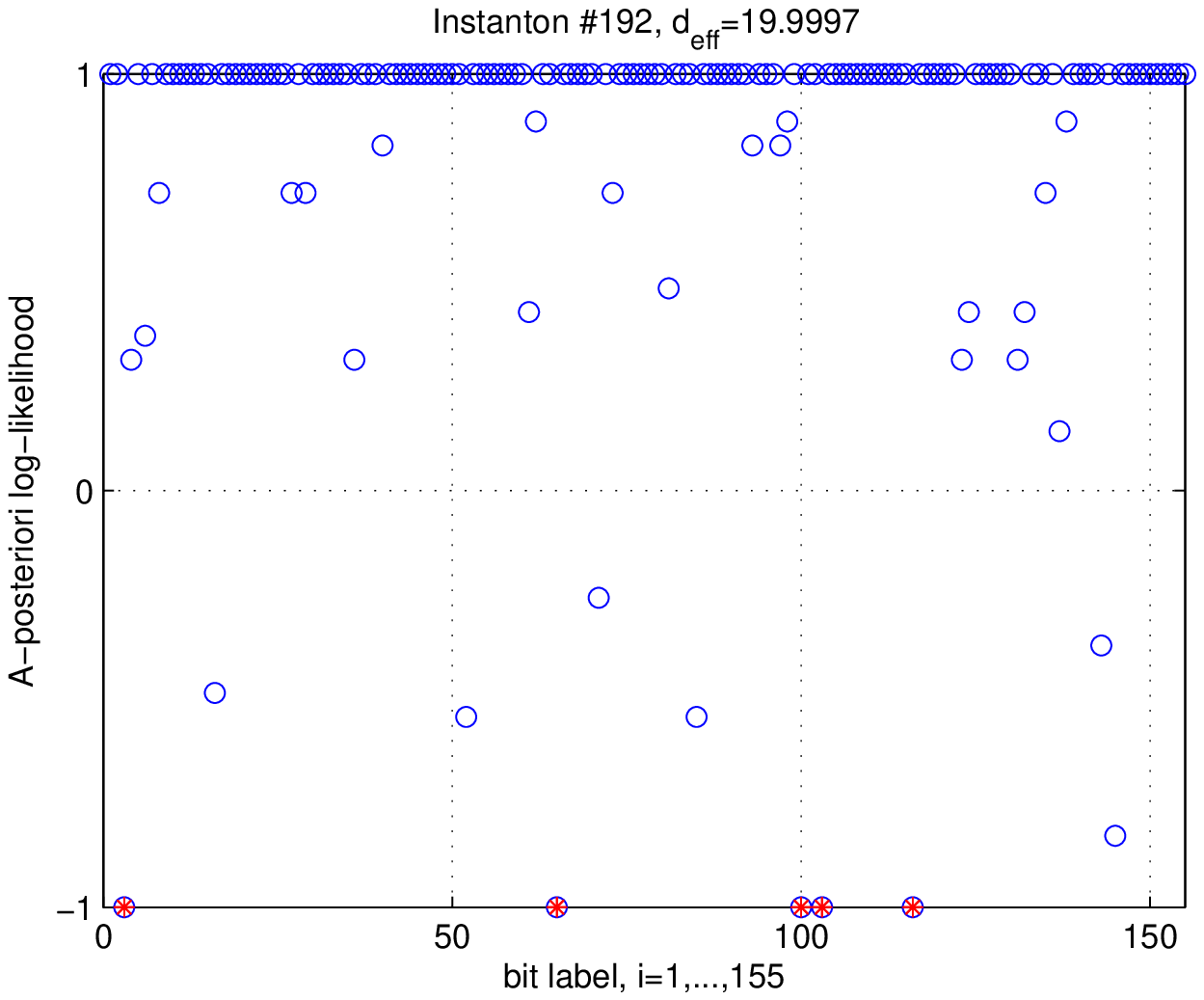}
\includegraphics{N192s.eps}}

\caption{The panels represent the results of LP decoding and
critical loop identification for $6$ representative instantons found
for the Tanner $(155,64,20)$ code performing over AWGN channel. All
instantons has an effective distance smaller than the Hamming
distance, $20$, of the code and the exact (ML) decoding would
correctly decode them to the correct codeword (all $+1$ for the test
shown in the Figure). The main Figures show dependence of the
a-posteriori log-likelihood on the bit label/position. Bits lying on
the critical loops are marked with red (filled circles). The
critical loops are also shown schematically in the subplots. Values
shown in the subplots next to the checks (squares) that connect
pairs of bits on the critical loop are related to the of the
corresponding triad contributions $\tilde{\mu}$ defined by
Eq.~(\ref{tildemu}). Loop-erasure decoding, with erasures applied
along the critical loop, corrects all the dangerous (instanton)
errors.} \label{N1}
\end{figure}

Fig.~\ref{N1} shows a representative set of instantons analyzed
using the described loop calculus tool. Critical loops are shown in
the corresponding subplots. The resulting critical loops are
typically $4-5$ bits long (with the girth of the code that counts
both bits and checks being $8$). However some configurations, like
instanton $\# 192$ shown in the right lower corner of
Fig.~(\ref{N1}), corresponds to a highly degenerate situation. The
instanton $\# 192$ shows three distinct single connected loops
giving all $r(\Gamma)=1$ contribution to the loop series
(\ref{Zseries}).

Obviously,  when an instanton produces a critical loop $\Gamma$ with
$r(\Gamma)=1$ one can guarantee that this is the largest non-bare
(non-BP) contribution to the loop series.  However  in all other
cases when $r(\Gamma)<1$, no guarantee can be given and we may not
exclude a possibility that some other general loop of a more complex
structure can provide a contribution to the loop series with higher
$r(\Gamma)$. However, we will show in Section \ref{sec:LP-er} that
knowledge of the $r(\Gamma)=O(1)$ critical contribution found along
the way explained above is sufficient for successful decoding of
these dangerous configurations.

One final remark of this Section concerns the value of magnetization
calculated for  degenerate critical loops, i.e. loops with
$r(\Gamma)=1$. Calculating a-posteriori log-likelihood
(magnetization) at a bit which belongs to the critical loop, one
finds that the first term in the numerator of Eq.~(\ref{m_full}) is
completely compensated by the only relevant of the other terms,
correspondent to $\Gamma_{a\to b}$ replaced just by the critical
loop $\Gamma$. Therefore, if only these two contributions are
accounted for the magnetization at the bit is exactly zero. This
suggests that one of the effects related to a critical loop is an
effective shift of log-likelihoods at the bits of the critical loop
in the direction opposite to the magnetization measured at the bit
by bare BP. We have also found that the bare BP
a-posteriori-log-likelihoods along the critical loop are always
aligned bit-wise with the corresponding log-likelihoods. This set of
observations will actually be explored in Section \ref{sec:LP-er} to
construct a simple modification of the LP decoding algorithm.

\section{Effective Free Energy Approach}
\label{sec:Eff}

Accounting for a single loop effect, when it is comparable to a bare
(BP) contribution, can be improved through an effective free energy
approach explained in this Section. This approach is akin to
degenerate Hartree-Fock variational approach used in quantum
mechanics (quantum chemistry) \cite{04DYK} in the case of a phase
space degeneracy.

It was shown in the previous Section that finding the BP gauge is
equivalent to optimizing (finding an extremum of) the functional
${\cal F}_0(\eta)=-\ln Z_0$, where $Z_0$ is  given by
Eq.~(\ref{Z0}). Therefore,  the $\eta$-gauge is fixed according to
the first term in the series Eq.~(\ref{Z-ready-gen}). Further terms
in $Z$, coming from other higher-order vertex ``corrections", were
calculated above with the BP gauge fixed. This resulted in
Eqs.~(\ref{gv_ser},\ref{Zseries}).

As we see from an example presented in the previous Section, some
number (or just one) of these corrections may be either comparable
(by absolute value) or just equal to the bare $Z_0$ contribution.
Any of this special $O(Z_0)$ contributions is associated with some
critical generalized loop. This situation, when one or more of these
critical loops emerge, is a potentially dangerous one, possibly
leading (as we observed) to the bare BP failure.

In this troublesome situation a plausible solution would be to modify the BP gauge conditions
Eq.~(\ref{extrZ0}) by
\begin{eqnarray}
\label{extrZ0Zloop} &&
\hspace{-1.4cm}\left.\frac{\delta\exp\left(-{\cal F}\right)}
{\delta\eta_{ab}}\right|_{\eta_{\rm eff}}\!\!\!=\!0,\ \ {\cal
F}\equiv -\ln\!\left(\!\!Z_0\!+\!\sum_{\Gamma}Z_{\Gamma}\!\right),
\end{eqnarray}
where $Z_{\Gamma}$ is the component of the full expression for $Z$ that corresponds to a critical
loop, $\Gamma$. In general Eq.~(\ref{extrZ0Zloop}) will be different from the standard BP equation,
and we can anticipates that the corrections due to the critical loops may cure the bare BP failure
in decoding.

From Eqs.~(\ref{extrZ0Zloop}) we derive the following set of
modified BP equations
\begin{eqnarray}
 && \hspace{-1cm} m_{a\to b}^{(*)}\!-\!m_{ab}^{(*)}\!=\!
 \sum_{\Gamma}\frac{
 \prod_{d\in \Gamma}\mu_{d;\Gamma}}
 {\prod_{(a'b')\in\Gamma}
 (1-(m_{a' b'}^{(*)})^2)} \ \delta m_{a\to b;\Gamma},\label{dm1}\\
 && \hspace{-1cm}\delta m_{a\to b}^{\Gamma}\!\!=\!\!\left\{\!\!\!\!
 \begin{array}{cc}
 \frac{1-(m_{ab}^{(*)})^2}
 {\mu_{b;\Gamma}}
 \langle\prod\limits_{c\in b;\Gamma}^{c\neq a}(\sigma_{bc}-m_{bc}^{(*)})\rangle_b,
 & (a b)\in \Gamma;\\ \hspace{-0.3cm}
 \frac{1-(m_{ab}^{(*)})^2}{\mu_{a;\Gamma}}
 \langle\hspace{-0.3cm}\prod\limits_{c\in a;\Gamma\ \& c=b}\hspace{-0.6cm}
 (\sigma_{ac}\!-\!m_{ac}^{(*)})\rangle_a,
 & \!\!\! a\!\in\!\Gamma, b\!\notin\!\Gamma;\\
 m_{ab}-m_{a\to b},
 & a\notin \Gamma;
 \end{array}
 \right.
 \label{dm2}\\
 &&\hspace{-1cm}
 \langle g({\bm\sigma}_a) \rangle_a\!\equiv\!\frac{\sum\limits_{{\bm\sigma}_a} g
 P_a({\bm\sigma}_a)}{\sum\limits_{{\bm\sigma}_a}
 P_a({\bm\sigma}_a)},\
 \mu_{a;\Gamma}\!\equiv\! \langle
 \prod\limits_{b\in\Gamma;a}\!
 (\sigma_{ab}\!-\!m_{ab}^{(*)})
 \rangle_a,
 \label{dm3}
\end{eqnarray}
where Eqs.(\ref{m1ab},\ref{m2ab}) defined $m_{ab}^{(*)}(\eta)$ and
$m_{a\to b}^{(*)}(\eta)$ and Eqs.~(\ref{dm1}) should all be taken at
$\eta\to\eta_{\rm eff}$ as the system of equations actually defines
$\eta_{\rm eff}$. Recast in terms of the beliefs Eqs.~(\ref{dm1})
forms a system of polynomial equations for beliefs, that become
linear only if the right hand sides of Eqs.~(\ref{dm1}) turn to
zero.

To find a-posteriori log-likelihoods within the degenerate approach
one calculates
\begin{eqnarray}
 m_{ab;{\rm eff}}\!\!=\!\!\frac{m_{a\to b}^{(*)}+\sum\limits_{\Gamma}\frac{
 \langle\sigma_{ab}\prod_{c\in
 a,\Gamma}(\sigma_{ac}-m_{ac}^{(*)})
 \rangle_a\prod_{d\in\Gamma}^{d\neq
 a}\mu_{d;\Gamma}}{\prod_{(a'b')\in\Gamma}
 (1-(m_{a' b'}^{(*)})^2)}}{
 1+\frac{
 \prod_{d\in \Gamma}\mu_{d;\Gamma}}
 {\prod_{(a'b')\in\Gamma}
 (1-(m_{a' b'}^{(*)})^2)}},
 \label{magEff}
\end{eqnarray}
where  $\eta$ is substituted by $\eta_{\rm eff}$ solving
Eqs.~(\ref{dm1}).

Note a couple of special cases. First of all,  if the graph consists
of a set of disconnected single loops, the modified BP equations
(\ref{dm1}) are simply reduced to the standard BP, with the rhs of
Eq.~(\ref{dm1}) replaced by zero. One consequence of this degeneracy
is that if one chooses ${\cal F}$ based on all the single-connected
loops contained in the degenerate model, the variational result
would just be exact. Second, if one considers contributions to
${\cal F}$ based on some number of single-connected critical loops,
$\Gamma$, the only terms on the rhs of Eqs.~(\ref{dm1}) that will
not result in exact zero for the bare BP solution will be associated
with $a\to b$ where $a\in \Gamma$ while $b\notin \Gamma$.

The modified free energy approach, described by equations
(\ref{dm1}) for renormalized gauges $\eta_{\rm eff}$,  promises
decoding benefit in the degenerate or close to the degenerate cases
when compared with the corresponding direct truncation of the loop
series given by Eqs.~(\ref{gv_ser}) or Eqs.~(\ref{Zseries}). The
approximation is also convenient as it keeps the same level of
complexity as the BP equations. This is in contrary to the direct
approach which requires summation according to Eq.~(\ref{m1}) over
many extended diagrams for getting renormalized values of
log-likelihoods at the bits that do not belong to the critical
loops.

Note that pretty much like in the case of bare BP, to define an
algorithm associated with Eqs.~(\ref{dm1}) one needs to introduce an
iterative algorithm based on it, and there is obviously some freedom
associated with the choice of discretization. (See \cite{06SCb} for
discussion of different discretization/iteration schemes in the
context of the bare BP equation.)

We expect that accounting for just one critical loop $\Gamma$ that corresponds to the largest value
of $r_\Gamma$ (calculated within the bare BP) will be already sufficient for a substantial
improvement of BP in the special cases when the bare BP fails. Summarizing, we arrive
at the following\\
\underline{\bf Loop-corrected BP algorithm}
\begin{itemize}
 \item{\bf 1.} Run bare BP algorithm. Terminate if BP succeeds
 (i.e. a valid code word is found in terms of marginal probabilities).
 \item{\bf 2.} If BP fails find the most relevant
loop $\Gamma$ that corresponds to the maximal (by absolute value)
amplitude $r_\Gamma$ in Eq.~(\ref{gv_ser}). A simple method for the
critical loop search introduced in the previous Section may be tried
first.
 \item{\bf 3.} Solve the modified-BP equations (\ref{dm1}) for the given $\Gamma$.
 Terminate if the improved-BP succeeds.
 \item{\bf 4.} Return to {\bf Step 2} with an improved $\Gamma$-loop
 selection. An additional loop found through an improved critical
 loop procedure can simply be added to the sum on the rhs of
 Eqs.~(\ref{dm1}).
\end{itemize}

In this manuscript we do not report any results of numerical simulations where the Loop-corrected
BP algorithm would be tested directly using a sample code. We postpone this important exercise for
future analysis. Instead, we use the result described in this Section as a motivation for an even
simpler heuristic approach detailed in the next Section.

\section{LP-erasure decoding}
\label{sec:LP-er}

As it was already discussed in the literature \cite{04VK,06CS} and
commented on in Section \ref{subsec:BFE}, LP decoding can be viewed
as a certain (large SNR) limit of BP decoding. It is not obvious,
however, that the BP-improved procedure outlined in the previous
Section can be rigorously transformed into a correction to LP
keeping a linear structure.

Our approach to this question will be heuristic, i.e. we simply
conjecture a plausible modification of the LP scheme based on the
algorithm formulated above for improved BP and then test this idea
using an example of the $(155,64,20)$ code.

On our way to proposing an improved LP decoding we first note that
the effective free energy approach keeps the same number of degrees
of freedom as the original bare BP.  Therefore, if we conjecture
that a modification of LP decoding should keep its linear structure
and thus the number of constraints, the only actual degree of
freedom left is in the log-likelihoods,  i.e. in their possible
modifications deduced from loop calculus. We further observe that
the modifications of the BP equations discussed in the previous
Section are actually well localized. Specifically, the rhs of
(\ref{dm1}) is non-zero for the $\eta$ variables associated with the
vertices that belong to the critical loop $\Gamma$ or are
immediately adjusted to it. Given that LP decoding is a special
limit of BP decoding one deduces from these observations that the
log-likelihoods should be renormalized just at the bits lying on the
critical loop. Furthermore,  taking into account the observation
reported in the last paragraph of Section \ref{sec:test}, we argue
that renormalization of log-likelihoods on the bits of the critical
loop should be directed against the bare log-likelihoods.

All this suggests the following LP-version of the loop-enhanced
algorithm:\\
\noindent \underline{\bf LP-erasure algorithm}
\begin{itemize}
 \item{\bf 1.} Run LP algorithm. Terminate if LP succeeds (i.e. a valid code word is found).
 \item{\bf 2.} If LP fails,  find the most relevant
loop $\Gamma$ that corresponds to the maximal amplitude $r(\Gamma)$ in the LP-version of
Eq.~(\ref{gv_ser}).
 \item{\bf 3.} Modify the log-likelihoods (factor-functions) along the loop $\Gamma$
introducing a shift towards zero,  i.e. introduce a complete or partial erasure of the
log-likelihoods at the bits. Run LP with modified log-likelihoods. Terminate if the modified LP
succeeds.
\item{\bf 4.} Return to {\bf Step 2} with an improved
selection principle for the critical loop.
\end{itemize}

Let us also mention that our loop-calculus based conjecture that the full or partial erasure of
certain log-likelihoods is beneficial for LP decoding is akin to the statement made in \cite{05FKV}
in regard to the positive effect of thresholding in LP decoding.

\subsection{$(155,64,20)$ test of the LP-erasure algorithm}
\label{subsec:test2}

Here we describe our numerical test of the LP-erasure algorithm. The
test is based on the analysis (described in Section \ref{sec:test})
of instantons, i.e. most probable out of a variety of dangerous
noise configurations that lead to LP-decoding failure. The $\sim 200
$ instantons found in \cite{06CS} for the $155,64,20)$ code have the
effective weight lower then the Hamming distance of the code,  thus
leading to the undesirable error-floor. We apply the triad method of
Section \ref{sec:test} to analyze all the low effective weight
instantons of the $(155,64,20)$ code. In spite of the fact that the
method did not guarantee that the critical loop was actually the one
with the largest (for given configuration of the noise) $r(\Gamma)$
we still choose to try it for the next, third, step of the
LP-erasure procedure: for the special marked bits we lowered the
original log-likelihood uniformly multiplying all the
log-likelihoods at the marked bits of the critical loop by a
positive number $\epsilon$ that is smaller than one.

The results of the test are remarkable.  We found out that all
instantons are actually corrected already with the roughest, i.e.
$\epsilon=0$, modification corresponding to the full erasure of the
information (log-likelihoods) along the critical loop. We  verified
that the noise configurations that are re-scaled instantons (of the
same structure but with effective distance larger then one of the
original instanton but smaller then the Hamming distance of the
code) are also corrected by the LP-erasure algorithm successfully.

Note that the instantons shown in Fig.~(\ref{N1}) are counted using
the all "+1" configuration as the original codeword primarily for
the purpose of the demonstration transparency. We did verify that
the LP-erasure algorithm is invariant with respect to a change in
the original codeword, i.e. that the LP-erasure algorithm corrects
instantons of the bare LP and their derived configurations if
counted using any other codeword as a reference point. In all our
tests (with the $\sim 200$ instantons) whenever LP-erasure decoded
to a codeword, the codeword was actually the right one.

The LP-erasure algorithm also shows an impressive robustness. The algorithm often forgives an
inaccurate definition of the critical loop.  For example, if one uses a very low threshold in
identifying the bits that can possibly enter the critical loop, the resulting loop can actually be
large and contain up to $20$ bits. By lowering or completely erasing log-likelihoods at all these
bits we often get the correct result with subsequent LP-decoding. However, in the rare cases when
this loose way to define a critical loop leads to a failure one just needs to tighten the threshold
and possibly use some additional thresholding (e.g. in the value of the a-posteriori-loglikelihood
that belong to the loop and also in the erasure coefficient $\epsilon$) till a codeword emerges.

\section{Conclusions and Discussions}
\label{sec:Con}

In this manuscript we presented a proof-of-concept test demonstrating the utility of the loop
calculus approach of \cite{06CCa,06CCb} to improve inference algorithms of the BP class in general,
and of BP decoding of LDPC codes in particular. The key observation, that enabled the reported
improvement in the decoding scheme, was emergence of a well-defined and relatively simple loop
correction for each BP-dangerous configuration of log-likelihoods. Identification of the critical
loop and further log-likelihood specific modification of the BP/LP algorithm has been suggested as
a cure for bare BP/LP failure. LP-erasure that is the simplest algorithm based on the critical loop
identification, was successfully tested using the $(155,64,20)$ code operated on the AWGN channel.
LP-erasure was able to correct all bare-LP-dangerous noise configurations related to the previously
found pseudo-codewords with the effective distance lower than the Hamming distance of the code.

This demonstration is clearly the first step in the highlighted direction where the next steps are
thought of as follows.   We plan to improve and continue testing the simple LP-erasure algorithm.
The major improvement required is an automatization of the critical loop identification scheme.
Further tests imply (a) direct comparison of LP-erasure with bare LP algorithms with the help of
Monte Carlo simulations of BER/FER, (b) working with other (longer) codes, (c) working with other
(e.g. correlated) channels. We will also be implementing all of the above using a more
sophisticated and also better justified Loop-corrected BP scheme outlined in Section \ref{sec:Eff}.

These studies will certainly benefit from using some other recent
developments in the field of BP/LP decoding, such as
\cite{06TS,06VK} on reducing complexity of LP-decoding, \cite{06SCb}
on accelerating the bare BP convergence , and \cite{06DW} which
suggests an alternative method of LP-decoding improvement.

\onecolumn

\twocolumn

\end{document}